\documentstyle{europhys}

\def\stars{\bigskip\centerline{***}\medskip}

\newif\ifboo \boofalse

\input epsf
\begin{document}
%
%
%
\euro{}{}{}{}
\Date{}
\shorttitle{}
%
%
%
\title{Exact Solutions of a Model for Granular Avalanches}

\author{Thorsten Emig, Philippe Claudin and Jean-Philippe Bouchaud}
\institute{Service de Physique de l'\'Etat Condens\'e, C.E. Saclay, Orme
des Merisiers, 91191 Gif-sur-Yvette Cedex, France}

%
%
\rec{}{}
%
%
%
\pacs{
\Pacs{45}{70.$-$n}{Granular systems}
\Pacs{45}{70.Ht}{Avalanches}
\Pacs{83}{50.Tq}{Wave propagation, shocks, fracture, and crack healing}
      }
\maketitle
%
%
%
\begin{abstract}
We present exact solutions of the non-linear {\sc bcre} model for granular
avalanches without diffusion. We assume a generic sandpile
profile consisting of two regions of constant but different slope.
Our solution is constructed in terms of characteristic curves from
which several novel predictions for experiments on avalanches are
deduced: Analytical results are given for the shock condition, shock
coordinates, universal quantities at the shock, slope relaxation at
large times, velocities of the active region and of the sandpile
profile.
\end{abstract}
%
%
%
%
%
\section{Introduction and Model}
The study of avalanches and surface flows in granular materials has
attracted much attention recently, both from a theoretical \cite{deG}
and an experimental point of view \cite{exp}. A simple model, thought
to capture some of the essential phenomena, has been proposed in
\cite{bcre.equa, bcre.hysteresis,bcre.triangular}. It is based on the
assumption that a strict separation between rolling grains and static
grains can be made. Coupled dynamical equations for these two species,
based on phenomenological arguments, can then be written. Calling $R$
the local density of rolling grains and $h$ the height of static
grains, the simplest form of the {\sc bcre} equations read:
\begin{eqnarray}
\label{eq:bcre1}
H_t & = & -\gamma R H_x, \\
R_t & = & R_x+\gamma R H_x,
\label{eq:bcre2}
\end{eqnarray}
where $H$ is the height of static grains, counted from the repose
slope of angle $\theta_r$: $h(t,x)=H(t,x)+x\tan(\theta_r)$ (the heap
is sloping upwards from left to right). In the above equations, the
units of lengths and time are chosen such that the (downhill) velocity
of grains is $v=1$, while $H$ and $R$ are counted in units of the
grains diameter. The term $\gamma R H_x$ describes the conversion of
static grains into rolling grains if $H_x >0$, or vice versa if
$H_x<0$. $\gamma$ is a grain collision frequency, typically of the
order of $100$ Hz.

Many important phenomenon are left out from the above description, and
can be included by adding more terms. For example, diffusion terms
(such as $D_1 R_{xx}$ or $D_2 H_{xx}$, describing, e.g. non local
dislodgement effects) will generically be present, and qualitatively
change the structure of the solutions \cite{bcre.triangular}. Another
aspect not described by the linear form of the conversion term above
is the expected saturation of rolling grains with time, rather than
the exponential growth predicted by Eq. (\ref{eq:bcre2}) for a
constant positive slope $H_x$.  Non linear saturation terms, as well
as a dependence of the velocity of the rolling grains on $R$, are thus
expected in general, and can lead to important differences with the
above equations \cite{pgdg.thick,pgdg.new}.

Recently, these equations has been studied by Mahadevan and Pomeau
({\sc mp}) \cite{mp}. They found a conservation law, which relates the
solutions $R(t,x)$ and $H(t,x)$ in a frame moving with the velocity of
the grains. From this law, they concluded that the {\sc bcre}
equations have characteristics that are straight lines, along which
both $R(t,x)$ and $H(t,x)$ are constant. Independently of the initial
profile $H_0(x)$, they found that a shock forms at time
$t_s=-1/(\gamma R_{0,\max}')$ with $R'_{0,\max}$ is the maximum (in
absolute value) of the initial gradient of rolling grains. Whereas our
exact solution fulfills the same conservation law, our results for the
characteristics and the shock time disagree with the results of {\sc
mp}.  As we will discuss below, the reason for this disagreement is
their implicit assumption of a very restrictive relation between the
initial profiles $R_0(x)$ and $H_0(x)$.

\section{Characteristic coordinates} 
The general basis of the method \cite{cf_book} we used to solve
Eqs. (\ref{eq:bcre1},\ref{eq:bcre2}) consists in a replacement of the
original equations by an equivalent system of four partial
differential equations for the functions $t$, $x$, $R$ and $H$, but
now considered as functions of new coordinates $\mu$ and $\nu$, which
will be defined below.\footnote{The theory used here is actually more
general and can be used in the presence of non-linear saturation terms
or for ripple models \cite{ripple-paper}.} These new equations will be
particularly simple inasmuch as each equation has derivatives with
respect to either $\mu$ or $\nu$, though the mapping between the
coordinates $(t,x)$ and $(\mu,\nu)$ will be in general complicated. To
define the characteristic coordinates $(\mu,\nu)$, we have to specify
first the characteristic curves of the system
(\ref{eq:bcre1},\ref{eq:bcre2}). For practical reasons, we introduce
new functions $u(t,x)=1-R(t,x)/\alpha$ and
$v(t,x)=(\alpha+x-R(t,x)-H(t,x))/\alpha$ instead of $H(t,x)$ and
$R(t,x)$.  For this new functions the differential expressions become
\begin{eqnarray}
\label{eq:dop1}
L_1[u,v] & = & -u_t - \gamma\alpha(1-u) u_x + v_t +\gamma\alpha (1-u)
v_x - \gamma(1-u)=0,\\
L_2[u,v] & = & u_t+[-1+\gamma\alpha(1-u)]u_x -\gamma\alpha(1-u)v_x
+\gamma(1-u)=0.
\label{eq:dop2}
\end{eqnarray}
Both operators $L_1$ and $L_2$ contain linear combinations of the type
$a u_t + b u_x$ of the derivatives of $u$ (and the same holds for
$v$). This combination means that $u$ is differentiated in the
direction given by the ratio $t/x=a/b$. Since the coefficients $a$ and
$b$ differ for $u$ and $v$ and also for $L_1$ and $L_2$, the functions
$u$ and $v$ are differentiated in each of the operators in different
directions in the $(t,x)$ plane.  Notice that the directions depend
also on $u$ itself, and therefore on the solution under consideration,
which is a typical feature of non-linear systems. As noted above, our
goal is to find equivalent differential equations of which each
contains derivatives in only one (local) direction corresponding to
one of the new coordinates $\mu$ and $\nu$. Therefore we take a linear
combination $L=\lambda_1 L_1 +
\lambda_2 L_2$ of the operators in Eqs. (\ref{eq:dop1},\ref{eq:dop2})
such that the derivatives of $u$ and $v$ in $L$ combine to derivatives
in the same direction, which is called a characteristic
direction. Moreover we assume that these local directions change
smoothly as functions of $t$ and $x$, and are given by the
tangential vectors $(t_\sigma(\sigma),x_\sigma(\sigma))$ of a smooth
path $(t(\sigma),x(\sigma))$ with $\sigma$ as parameter. Considering
the functions $u$ and $v$ along this path, they depend only on
$\sigma$ and we have, e.g., $u_\sigma=u_t t_\sigma + u_x
x_\sigma$. Using these conditions, we obtain four
homogeneous linear equations for the coefficients $\lambda_1$ and
$\lambda_2$ with coefficients depending on $t$, $x$, $u$, $v$ and
their derivatives with respect to $\sigma$. For non-trivial solutions
all possible determinants of the matrix of these coefficients have to
vanish, leading to three independent equations or characteristic
relations ({\sc cr}). The first one can be written as a quadratic equation
for the local direction $\zeta=x_\sigma/t_\sigma$ of
differentiation, the solution of which are: 
$\zeta_+=-1$ and $\zeta_-=\gamma\alpha(1-u)$.
Now, for a fixed solution $u$, the equations $dx/dt=\zeta_+$ and
$dx/dt=\zeta_-$ are ordinary differential equations, which define two
families of paths with the starting position $x_0$ at $t=0$ as
parameter. These families of paths are the characteristics $C_+$ and
$C_-$ of the system (\ref{eq:bcre1},\ref{eq:bcre2}). From a physical
point of view, they are simply the paths along which $R(x,t)$ 
($\zeta_+$) and $H(t,x)$ ($\zeta_-$) evolves with time.

The new curved coordinate frame $(\mu,\nu)$ is now defined such that
the two one-parametric families of characteristics are mapped by the
coordinate transformation on an usual Cartesian coordinate frame in
the $(\mu,\nu)$-plane, i.e., along the characteristics the coordinate
functions $\mu(t,x)$ and $\nu(t,x)$, respectively, are constant. Here
we have chosen to map the line $t=0$ on the line given by $\mu=-\nu$.
In terms of the new coordinates we find
\begin{equation}
\label{eq:ch1}
x_\nu+t_\nu=0, \quad x_\mu-\gamma\alpha(1-u)t_\mu=0.
\end{equation}
Now we make use of another {\sc cr}, which evaluated along $C_+$ and $C_-$
by identifying $\sigma$ with $\nu$ and $\mu$, respectively, yields the
conditions
\begin{equation}
\label{eq:ch2}
u_\nu+\gamma\alpha(1-u)v_\nu+\gamma(1-u)t_\nu=0, \quad
u_\mu-v_\mu+\gamma(1-u)t_\mu=0.
\end{equation}
These equations together with the Eqs. (\ref{eq:ch1}) form the desired
set of four equations mentioned before. Every solution of this new
system satisfies the original Eqs. (\ref{eq:bcre1},\ref{eq:bcre2}),
since the Jacobian $t_\nu x_\mu - t_\mu x_\nu \sim 1+\gamma
R(\mu,\nu)$ of the coordinate map does not vanish due to $\gamma
R(\mu,\nu)>0$.  

\section{General solution}  Before we can construct a solution to the equivalent
system (\ref{eq:ch1},\ref{eq:ch2}), we have to specify initial data
along the line $\mu=-\nu$ corresponding to $t=0$. We choose an general profile
$H_0(x)$, perturbed at $t=0$ by a uniform `rain' of rolling grains: $R_0(x)=\alpha$. In terms of the new coordinates, the initial conditions become
$t_0(\mu)=0$, $x_0(\mu)=-\mu$, $u_0(\mu)=0$, $
v_0(\mu)=-(\mu+H_0(-\mu))/\alpha$.
By introducing the function
$\Delta(\mu,\nu)=-1-\gamma\alpha(1-u(\mu,\nu))$, one can show that the
problem of solving the system given by
Eqs. (\ref{eq:ch1},\ref{eq:ch2}) can be reduced to the task of finding
a solution to the equation
$\Delta_\nu = \gamma H'_0(\nu)(1+1/\Delta)$,
with initial condition $\Delta(\mu,-\mu)=-1-\gamma\alpha$. The
solution of this equation can be simply expressed in terms of the
so-called Lambert function $W$ \cite{Lambert}:
\begin{equation}
\label{eq:delta}
\Delta(\mu,\nu)=-1-W\left\{\alpha\gamma \exp[\alpha\gamma + \gamma(H_0(-\mu)-
H_0(\nu)) ]\right\}.
\end{equation}
With this solution at hand, the solution to the system
(\ref{eq:ch1},\ref{eq:ch2}) is determined by
\begin{eqnarray}
t(\mu,\nu)=\int_{-\nu}^\mu \frac{ds}{\Delta(s,\nu)} 
= -\mu-\nu+\int_{-\nu}^\mu \frac{\Delta_\mu(s,\nu)ds}{\gamma H'_0(-s)} 
&,& \quad
x(\mu,\nu)=-\mu-t(\mu,\nu) \nonumber\\
R(\mu,\nu)=-\frac{1+\Delta(\mu,\nu)}{\gamma} &,& \quad 
H(\mu,\nu)=H_0(\nu),
\label{gen_solution}
\end{eqnarray}
where we have expressed already the original fields $R(\mu,\nu)$ and
$H(\mu,\nu)$ in terms of the functions $u$ and $v$. To get the fields
as functions of $t$ and $x$, one has to invert the coordinate
map. This can be done by using $\mu(t,x)=-t-x$ and integrating the
equation for $t(\mu,\nu)$ to obtain also $\nu(t,x)$. As announced
before, the height profile $H(t,x)=H_0(\nu(t,x))$ turns out to be
constant along the characteristics $C_-$.

\section{Generic shape for $H(t,x)$} 
In the following we will consider a situation which is generic for
sandpile surfaces. Suppose that one starts with a sandpile profile,
which consists of two regions with constant but different slopes
matching with a kink at $x=0$, and again with a constant amount of
rolling grains. The slopes may be either larger or smaller than the
angle of repose $\theta_r$. If we denote the slope to the right (left)
by $\theta_r+\theta_>$ ($\theta_r+\theta_<$), we have $H_0(x)=\theta_>
x$ for $x>0$ and $H_0(x)=\theta_< x$ for $x<0$. In the case of a
piecewise constant $H_0'(x)$ one can integrate the equation for
$t(\mu,\nu)$ easily as can be seen from Eq. (\ref{gen_solution}).  The
structure of Eq. (\ref{eq:delta}) suggests to distinguish between
three regions given by $\mu>0$,$\nu<0$ (I), $\mu,\nu<0$ (II) and
$\mu<0,\nu>0$ (III). \footnote{The region where $\mu,\nu>0$ turns out
to be mapped on the half space with $t<0$ and is therefore not of
physical interest.} In regions I and III one can find the explicit
expression $\nu(t,x)=x + \frac{\alpha}{\theta} (1 - e^{\gamma\theta
t})$ with $\theta=\theta_<$ (I) or $\theta=\theta_>$ (III), i.e., the
characteristics $C_-$ are in these regions simple exponential
curves. As a consequence, no shocks can appear in these two regions
and the corresponding solutions are particularly simple:
\begin{equation}
\label{eq:RHinreg13}
R_{I(III)}(t,x) = \alpha e^{\gamma\theta_{<(>)} t}, \quad 
H_{I(III)}(t,x) = H_0(x) +\alpha - R_{I(III)}(t,x).
\end{equation}
The boundaries of the regions I and III in real space $(t,x)$ are
given by the conditions $x<-t$ and $x>x_1(t)=\frac{\alpha}{\theta_>}
(e^{\gamma\theta_> t}-1)$ corresponding to the $\mu=0$ and $\nu=0$
characteristics, respectively, see Fig. \ref{fig1}. The boundary for
region I has an obvious physical meaning: The information that there
is a kink at $x=0$ can only propagate to the left with the velocity of
the moving grains, which is $1$ in our rescaled units. Moreover, it is
important to note that the `uphill' velocity with which the kink moves
is only equal to $\gamma \alpha$ at small times, before growing
exponentially.  As discussed in the introduction, this growth
eventually saturates, as does the value $R$, or else the
characteristic $C_-$ quickly reaches the edge of the pile.

The range of $x$ in between the above two regions corresponds to the
intermediate region II. Within this range one can obtain only an
implicit solution for the coordinate map $\nu(t,x)$. It reads
\begin{equation}
\label{eq:nuinreg2}
\nu = x + \frac{1}{\gamma} \left [
\frac{\Delta(-x-t,\nu) - \Delta(0,\nu)}{\theta_>} +
\frac{\Delta(0,\nu) + 1 + \alpha\gamma}{\theta_<} \right ],
\end{equation}
where $\Delta(\mu,\nu)=-1-W\left\{\alpha\gamma \exp[\alpha\gamma -
\gamma(\theta_> \mu + \theta_< \nu)]\right\}$ as follows from 
Eq. (\ref{eq:delta}). The shape of $R(t,x)$ and $H(t,x)$ can be
obtained directly from the last two equations of (\ref{gen_solution}).
In general, Eq. (\ref{eq:nuinreg2}) has to be solved numerically
although several results can be obtained in an analytic way. It turns
out that the solutions of Eq. (\ref{eq:nuinreg2}) fall into two
qualitatively different classes, according to the values of
$\beta=\theta_>/\theta_<$ and $\theta_<$: for $\beta > 1-\alpha\gamma$
or $\theta_< <0$, both $R(t,x)$ and $H(t,x)$ remain continuous for all
times, while for $\beta < 1-\alpha\gamma$ and $\theta_< >0$, the
solutions develop a discontinuity in $R(t,x)$ and $H(t,x)$ beyond a
finite shock time $t_s$.  This must be contrasted with {\sc mp}, since
in the present case $R_0(x)=\alpha$, they predict that shocks are
absent for all times.

\section{Examples}
\begin{figure}[phtb]
\begin{center}
\leavevmode
\epsfxsize=0.49\linewidth
\epsfbox{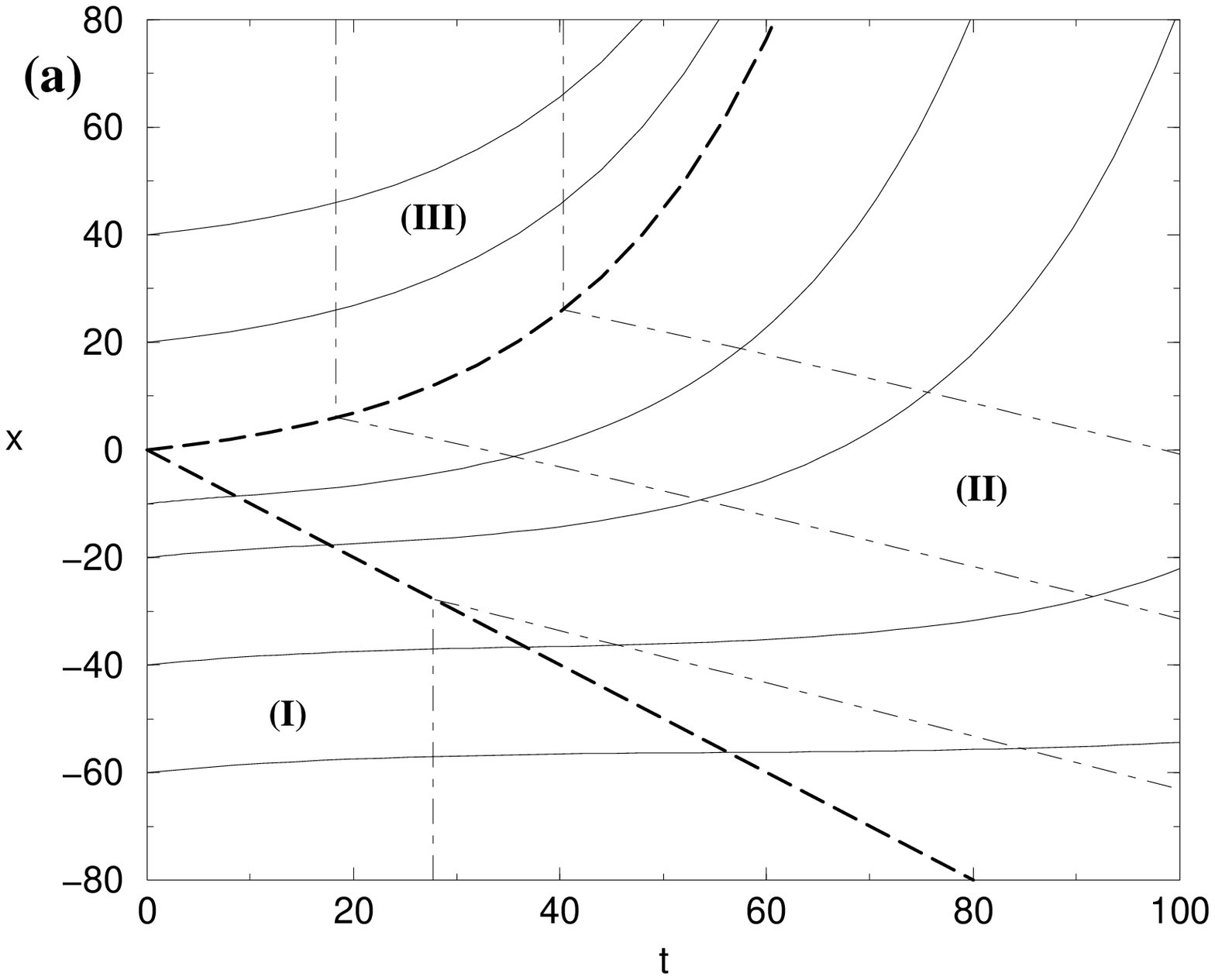}
\hfill
\epsfxsize=0.49\linewidth
\epsfbox{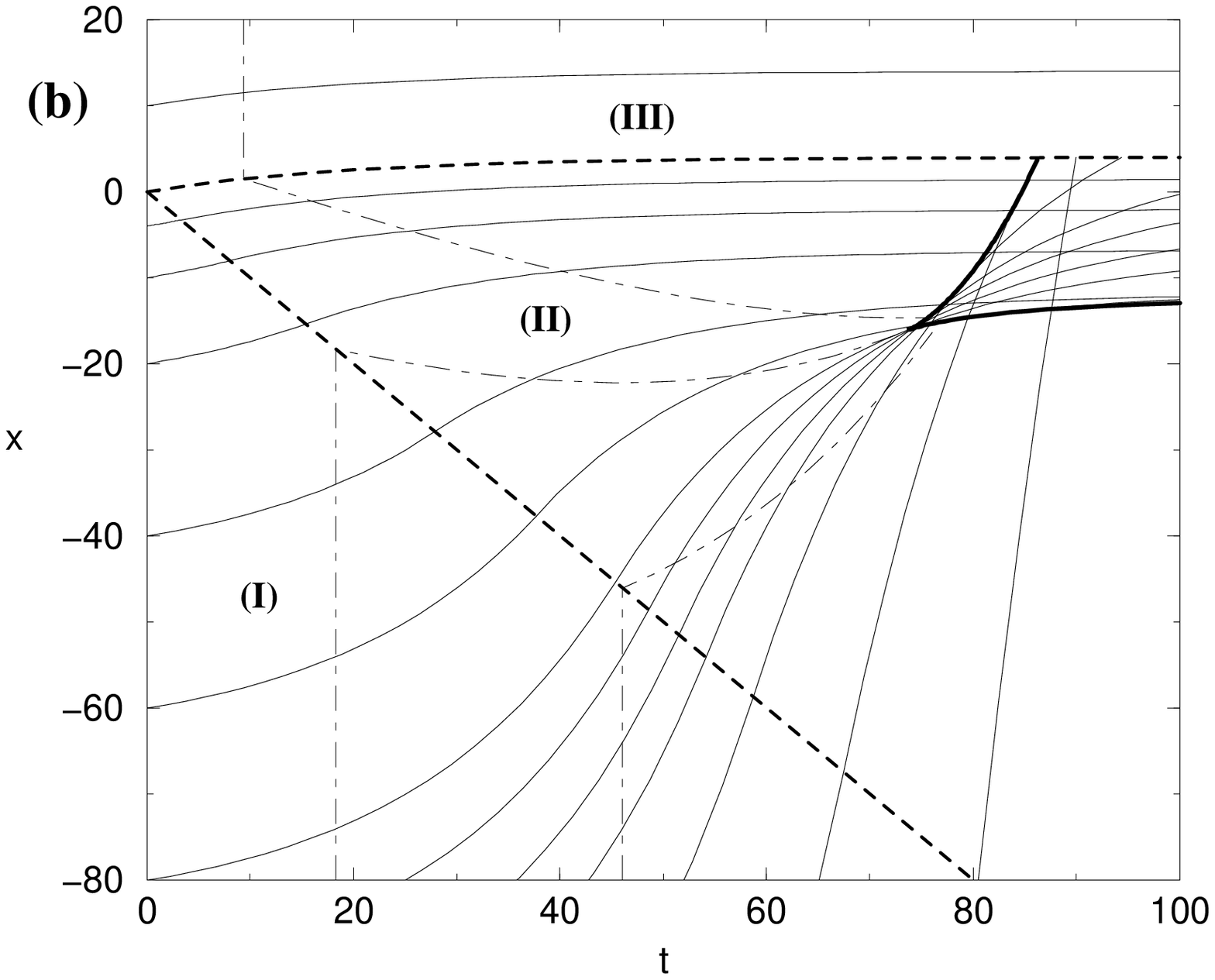}
\end{center}
\caption{Characteristics $C_-$ for the cases (a) $\theta_>=0.1$,
$\theta_<=-0.1$ and (b) $\theta_>=-0.1$, $\theta_<=0.1$. For both
cases we have taken $\gamma=0.5$, $\alpha=0.4$. The dashed lines
represent the boundaries between the different regions I, II and III
explained in the text. The bold line in (b) is the envelop of the
characteristics. It presents a kink at the shock position, where the
characteristics cross for the first time. Along the dot-dashed lines
$R(t,x)$ is constant.}
\label{fig1}
\end{figure}

The characteristics resulting from numerical solutions of
Eq. (\ref{eq:nuinreg2}) have been plotted in Fig. \ref{fig1}. The left
part of this figure has been obtained for $\theta_> > 0$ and $\theta_<
< 0$, corresponding to $\beta<0$.  In this case, the characteristics
are more and more `diluted' as time increases, and therefore never
cross -- no shock.  In the limit of large times, the argument of the
Lambert $W$ function becomes very large. Using the first two terms of
the asymptotic expansion of $W$
\cite{Lambert} we get $\nu(t,x)=[-\beta t+\ln( x + \frac{\beta t} {\beta - 1}
)/(\gamma\theta_<)]/(\beta-1)$.  The corresponding expression for
$R(t,x)$ and $H(t,x)$ can be obtained from Eq. (\ref{gen_solution}). A
particularly interesting quantity to look at is the local slope at,
say, $x=0$. In this limit the slope is negative and decays with time
as $H_x(t,x=0)=1/(\gamma\beta t)$.\footnote{Note that this $t^{-1}$
relaxation of the slope has also been obtained in \cite{HK} within a
very different model.} It means that the `true' slope $h_x$ actually
relaxes to the angle of repose $\theta_r$ for very large time.  If $L$
is the size of experimental system, then $C_-$ reaches the boundary of
the system at a time $t^*$ such that $L \approx
\frac{\alpha}{\theta_>}e^{\gamma \theta_> t^*}$. One should therefore
measure a final slope $h_x \approx \theta_r +
\theta_</\ln ( \theta_> L /\alpha )$ {\it smaller} than the repose
angle.  This result is consistent with the qualitative discussion of
Boutreux and de Gennes for a similar situation \cite{pgdg.sinai}.

Another experimentally important quantity is the velocity $v_R$ of the
``active'' region. Following \cite{bcre.triangular}, this region can
be defined by the condition $R(t,x)>R_{\rm min}$, where $R_{\rm min}$
is a small threshold. $v_R$ is then given by the slope of the curves
of constant $R(t,x)$, which tends to a constant in the large $t$ limit
as can be seen in Fig. \ref{fig1}(a). The asymptotic analysis yields
$v_R=\beta/(1-\beta)$. Since $\beta<0$, $-1<v_R<0$, and the avalanche
proceeds {\it downhill}, but slower than the grains themselves. This
is an effect of the non-linear term in the {\sc bcre} equations since
the linearized theory yields $v_R=-1$ \cite{bcre.triangular}.

\begin{figure}[phtb]
\begin{center}
\leavevmode
\epsfxsize=0.49\linewidth
\epsfbox{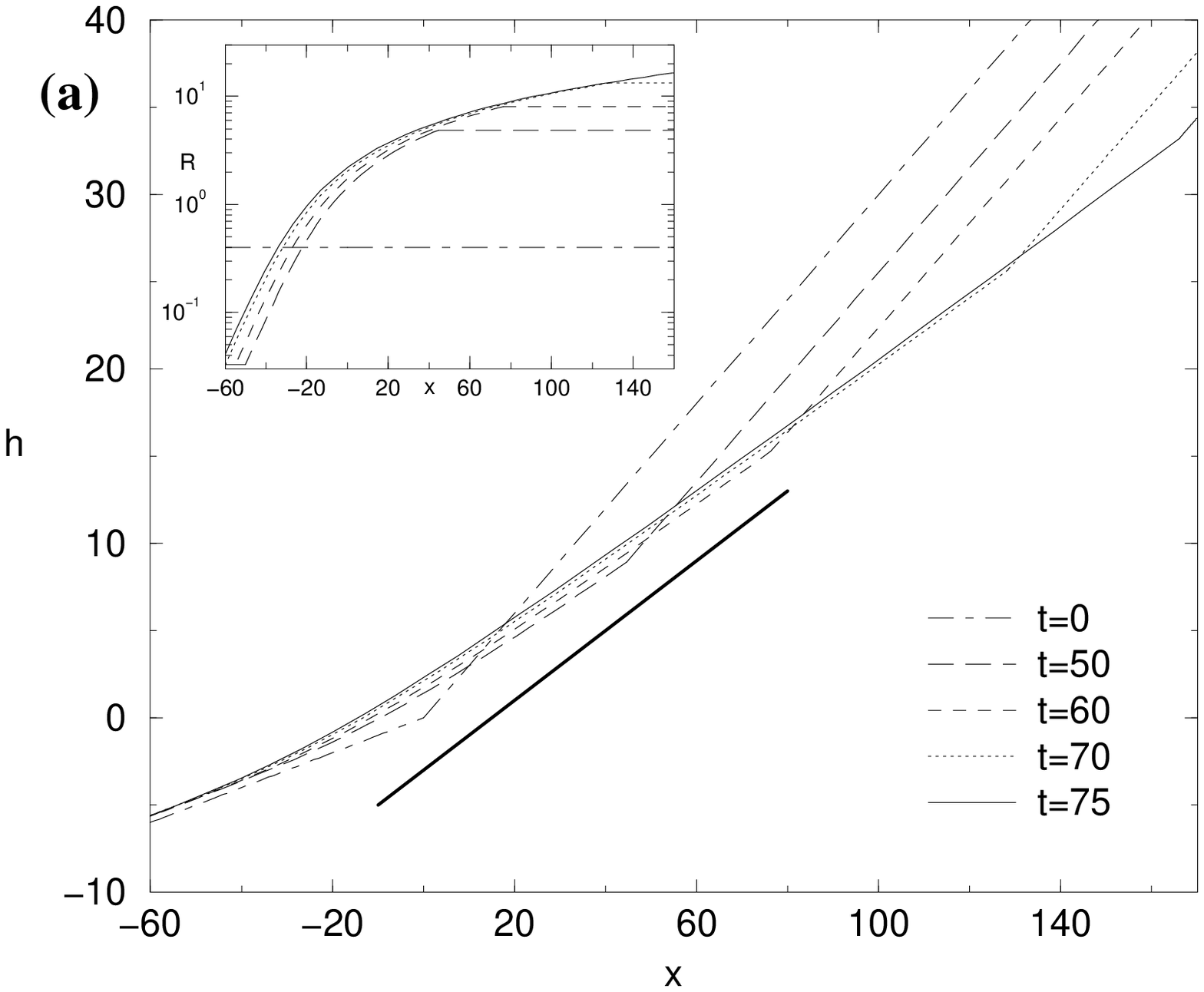}
\hfill
\epsfxsize=0.49\linewidth
\epsfbox{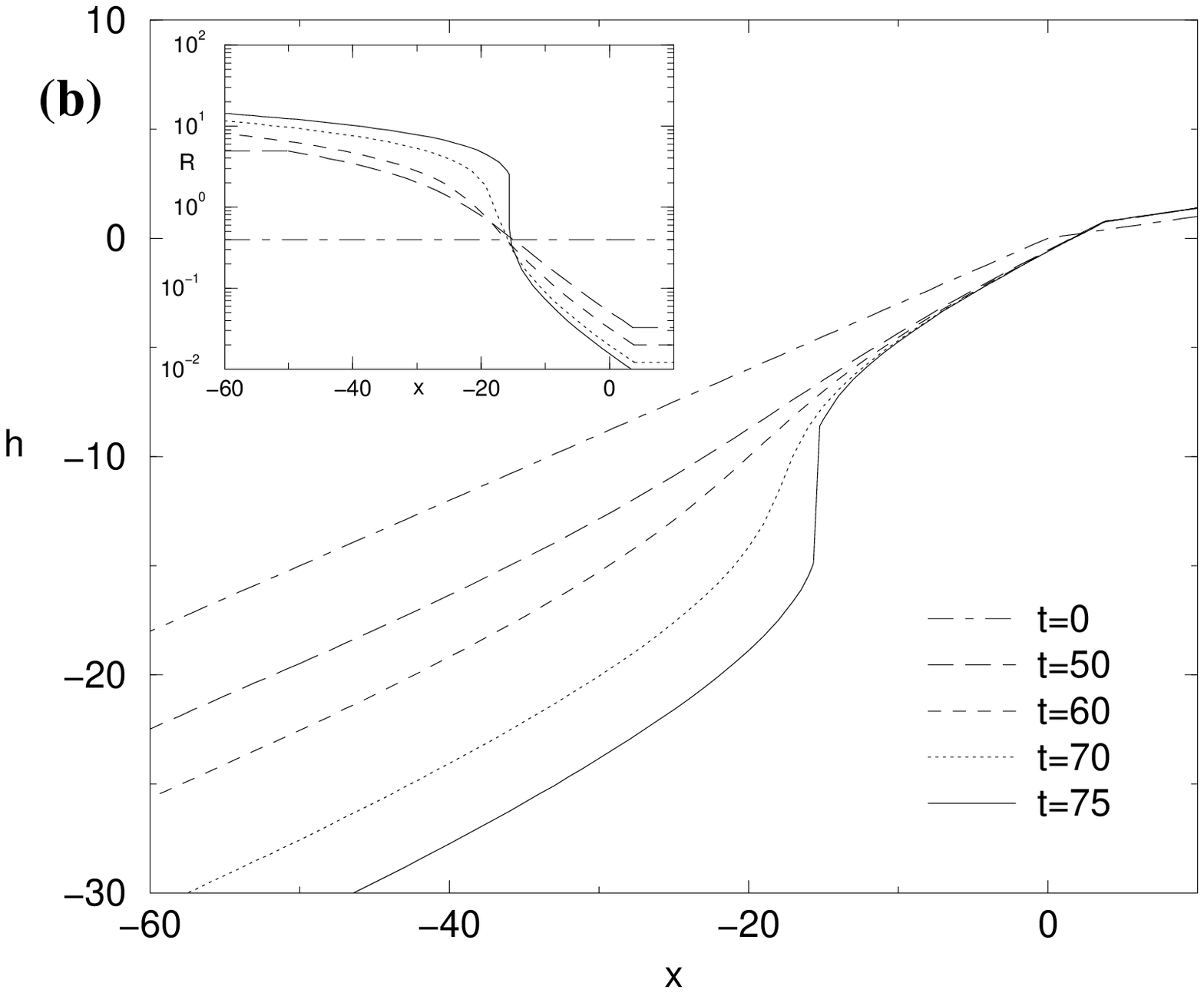}
\end{center}
\caption{Total height profiles $h(t,x)$ for the two cases of Fig. \ref{fig1}.
The bold line in (a) shows the critical slope, which is chosen here as
$\theta_r = 0.2$. In (b) the shock occurs at $t_s=73.78$,
$x_s=-16.05$. Insets: Corresponding evolution of the amount of moving
grains $R(t,x)$. For $t=75$ the solutions $h(t,x)$ and $R(t,x)$ are
not single valued for $-15.61<x<-15.20$ since each point within the
range bounded by the envelop is covered three times by a
characteristic $C_-$. As orientation guide, the limiting values at
both boundaries are connected here by straight lines.}
\label{fig2}
\end{figure}

The situation where $\theta_> < 0$ and $\theta_< > 0$ is qualitatively
different. In this case, the characteristics cross at some finite
time: a shock occurs -- see Fig. \ref{fig1}(b). A crossing point of
two characteristics means indeed that at this point, two different
values of $R$ (or $H$) are possible and these functions then become
discontinuous. Strictly speaking, the Eqs. (\ref{eq:bcre1},\ref{eq:bcre2})
are no longer valid, and the diffusion terms left of from the analysis
become important to smooth out this discontinuity. In Fig. \ref{fig2}, we plotted snapshots of the $h$ and $R$ profiles at different times, for both 
situations (with and without the occurrence of a shock).

One can calculate the time $t_s$ and location $x_s$ at which
the shock occurs. For that purpose, let us introduce the envelop of
the characteristic curves $x(t,\nu)$, where $\nu$ is a label. The
envelop can be represented in a parametric way as $\left ( t_e(\nu),
x_e(\nu)
\right )$. It has the property that for each of its points exists a
characteristic, which touches it tangentially.  It has then to fulfill
the conditions $x(t_e(\nu),\nu) = x_e(\nu)$, $x_\nu ( t_e(\nu),
\nu ) = 0$. After some calculations, one can find the {\em explicit} expression
for the envelop,
\begin{eqnarray}
\label{eq:xe}
x_e(\nu) & = & \nu - \frac{1}{\gamma\theta_<} \left [
1 + \alpha\gamma + \Delta(0,\nu) \left ( 1 +
\frac{1}{1-\beta} \right ) \right ] \\
\label{eq:te}
t_e(\nu) & = & - x_e(\nu) + \frac{\nu}{\beta} -
\frac{1}{\gamma\theta_>} \left [ \alpha\gamma + \ln (\alpha\gamma)
+ 1 + \frac{\Delta(0,\nu)}{1-\beta} 
- \ln \left ( -1 - \frac{\Delta(0,\nu)}{1-\beta}
\right ) \right ].
\end{eqnarray}
This envelop has two branches, separated by a kink, see
Fig. \ref{fig1}(b), given by $\nu=\nu_s = [\alpha\gamma + \ln
(\alpha\gamma)- 1 + \beta - \ln \left ( 1 - \beta \right
)]/(\gamma\theta_<)$. Whereas the upper branch is parameterized by
$-\infty < \nu < \nu_s$, the lower one corresponds to $\nu_s < \nu <
\nu_c=[\beta+\alpha\gamma+\ln(-\alpha\gamma/\beta)]/(\gamma\theta_<)$.
The resulting shock coordinates are
\begin{equation}
\label{eq:xsts}
x_s  =  \frac{1}{\gamma\theta_<} \left [\ln(\alpha\gamma)-
\ln(1-\beta)+1+\frac{1}{1-\beta}\right], \quad
t_s  = \frac{1}{\gamma\theta_<}\left[\left(1-\frac{2}{\beta}
\right)\ln\left(1-\beta\right)-\ln(\alpha\gamma)\right].
\end{equation}
The condition that ($t_s,x_s$) has to be located inside region II
leads to the boundary between the classes with and without shock as
mentioned above. At the shock position, the amount of moving grains
is {\it universal} (independent of the initial value $\alpha$), and
given by $R_s=1/(\gamma(1-\beta))$, while $H_s=\theta_<
\nu_s$. Since typically $v \sim \gamma d$ with $d$ the grain diameter, 
we have in our rescaled units $\gamma\sim 1$ showing that due to $R_s
\stackrel{<}{\sim} 1$ non linear saturation terms can be neglected at 
the shock if $\beta \stackrel{<}{\sim} -1$.  The lower branch of the
envelop saturates for large $t$ exponentially fast with a
characteristic time $1/(\gamma\theta_>)$ at
$x_\infty=[1+\ln(-\alpha\gamma/\beta)]/(\gamma\theta_<)$, which is
always larger than $x_s$. This means that the shock stops propagating
upwards. A large time expansion in the shock free range
$-t<x<x_\infty$ gives, taking the two leading terms of $W$, $\nu(t,x)=
-(\alpha/\theta_<)\exp[\gamma\theta_<t - (\theta_</\alpha) x
e^{-\gamma\theta_<t}]$. Thus the slope is non monotonous within this
range: after increasing for small times it relaxes again to the initial
value $\theta_<$ as $H_x(t,x)=\theta_< \exp[-(\theta_</\alpha) x 
e^{-\gamma \theta_< t}]$.

\section{Discussion}
Let us summarize the major results of this paper, which could be
explored experimentally. Starting from an initial profile made up of
two different slopes, we find that shocks can occur after a finite
time, depending on the value of the two slopes and the initial density
of rolling grains. When shocks are absent, we find that the evolution
surface profile is characterized by different velocities: the kink
moves upwards with a velocity of the order of $\alpha \gamma$ for
early times, while the edge of the ``active'' region moves downwards
at a velocity which only depends on the initial slopes, and is smaller
than the velocity of the grains.  The final slope is shown to be the
angle of repose; however, for finite size systems, one expects the
final slope to be smaller by an amount which varies as $1/\ln L$. When
a shock appears, we predict the time and position of this shock, as
well as the density of rolling grains there, which takes a universal
value. The shock is found to stop progressing upwards.

Our results are in disagreement with those of {\sc mp}. For the
situation considered here, they predict that the initial profile is
rigidly shifted along straight characteristics. Therefore, for
example, the final slope would be given by $H_x(t,x=0)=\theta_<$,
which is completely different from our prediction of a decaying slope.
The reason for this discrepancy comes from their implicit assumption
that $R_0(x)+H_0(x)+\ln(R_0(x))/\gamma={\rm const.}$, which does not
hold in the cases considered here.

The method presented here can be extended to more general
situations. For example, each profile $H_0(x)$ can be approximated by
a piecewise linear function.  Therefore, our analysis can be used to
obtain analytical results for more complicated situations as, e.g.,
bumps or sinusoidal shapes. Another interesting situation is the case
where $R_0(x)$ is localized in space. Applications of this method to
the problem of ripple formation are under way.

Two important physical phenomena have been neglected: diffusion terms,
which are expected to be important in the presence of shocks or in the
case of a localized initial $R_0(x)$ (see \cite{bcre.triangular}), and
non-linear effects, which lead to a saturation of the static/rolling
grains conversion term.  A simple way to account for the latter effect
is to replace the characteristics by straight lines of velocity
$\gamma R_\infty$ as soon as $R=R_\infty$. The influence of a
dependence of the velocity of grains on their density would also be
worth investigating \cite{pgdg.new}.

\stars

This research was partly supported by the Deutsche Forschungsgemeinschaft
(DFG) under grant EM70/1-1.

%

\end{document}